\documentclass[showpacs,aps,amsmath,letterpaper,nofootinbib,reprint,showkeys,superscriptaddress,pra]{revtex4-2}
\usepackage{dcolumn}    % Align table columns on 
\usepackage{bm}         % bold math
\usepackage{ifpdf}
\usepackage{amssymb,lineno,amsfonts}
\usepackage{graphicx}   % Include figure files
\usepackage{bbm}
\usepackage{mathrsfs}
\usepackage{mathtools}
\usepackage{upgreek}

\usepackage{epstopdf}
\usepackage{setspace}
\usepackage{hyperref}
\usepackage{cleveref}
\usepackage{bbold}
\usepackage[matrix,frame,arrow]{xypic}
\usepackage{float}
\usepackage{soul}
\usepackage{tikz}
\usepackage{natbib}
\usepackage{color} 
\usepackage{subfigure}
\usepackage{bbold}
\usepackage{tikz}
\usepackage{amsthm}
\usepackage{academicons}
\usepackage[english]{babel}

\definecolor{med-blue}{RGB}{25,25,112} 
\hypersetup{colorlinks, linkcolor={blue},citecolor={blue}, urlcolor={blue}}

\newcommand{\Eqref}[1]{(\ref{#1})}

\newcommand{\expo}[1]{\mathrm{e}^{#1}}
\newcommand{\im}{\mathrm{i}}
\newcommand{\outpr}[2]{\vert{#1}\rangle\langle{#2}\vert}
\newcommand{\proj}[1]{\vert{#1}\rangle \langle{#1}\vert}
\newcommand{\Ket}[1]{\vert{#1}\rangle}
\newcommand{\Bra}[1]{\langle{#1}\vert}

\newcommand{\dnorm}[1]{\vert\vert{#1}\vert\vert}

\usepackage{orcidlink}

\begin{document}
\title{ Enhanced non-macrorealism: Extreme violations of Leggett-Garg inequalities for a system evolving under superposition of unitaries}
\author{Arijit Chatterjee$^{\ddagger}$ \orcidlink{0009-0000-8191-1882}}
\email{arijitchattopadhyay01@gmail.com}
\affiliation{Department of Physics and NMR Research Center,\\
Indian Institute of Science Education and Research, Pune 411008, India}
\author{H.~S.~Karthik$^{\ddagger}$}
\email{hsk1729@gmail.com}
\affiliation{International Centre for Theory of Quantum Technologies, University of Gdansk, 80-308 Gdansk, Poland}
\author{T. S. Mahesh}
\email{tsmahesh@gmail.com}
\affiliation{Department of Physics and NMR Research Center,\\
Indian Institute of Science Education and Research, Pune 411008, India}
\author{A. R. Usha Devi}
\email{ushadevi@bub.ernet.in}
\affiliation{Department of Physics, Bangalore University, Bengaluru-560 056, India}
\def\thefootnote{$\ddagger$}\footnotetext{These authors contributed equally to this work}\def\thefootnote{\arabic{footnote}}

\begin{abstract}
{ Quantum theory contravenes classical macrorealism by  allowing a system to be in a superposition of two or more physically distinct states, producing physical consequences radically different from that of classical physics. We show that a system , upon subjecting to transform under superposition of unitary operators, exhibits enhanced non-macrorealistic feature -  as quantified by violation of the Leggett-Garg inequality (LGI) beyond the temporal Tsirelson bound. Moreover, this superposition of unitaries also provides robustness against decoherence by allowing the system to violate LGI and thereby retain its non-macrorealistic behavior for a strikingly longer duration. Using an NMR register, we experimentally demonstrate the superposition of unitaries with the help of an ancillary qubit and verify these theoretical predictions.}
\end{abstract}

\maketitle	

\emph{Introduction---}The twilight zone between classical and quantum worlds \cite{RevModPhys.75.715,zurek1993preferred,zurek2014quantum} has been under perpetual investigation since the dawn of quantum mechanics \cite{PhysRev.47.777, PhysicsPhysiqueFizika.1.195,feynman2000theory}.   The primary foundational difference comes from quantum theory's radical take on `macrorealism' and `locality'- two of the most cherished notions of classical physics \cite{PhysRev.47.777, PhysicsPhysiqueFizika.1.195, PhysRevLett.49.1804}. The word `macrorealism' means that a system, capable of being in two or more physically distinct states, will always be in any one of them. This stands in a strong contradiction with quantum theory, which allows the system to be in a superposition of physically distinct states. Superposition of quantum states also permits multiple systems to be in an entangled state \cite{PhysRev.47.777,RevModPhys.81.865} where the instantaneous behavior of any one system depends on the other systems even though large distances separate them. This leads to serious debates about the notion of `locality'\cite{pan2000experimental, PhysRevLett.49.1804}.

The quantum  behaviour of a system is, therefore, primarily accredited to its capability of being in superposition of physically distinct states. Classicality emerges when the system losses the superpositions and turns into an incoherent mixture through decoherence.\cite{hornberger2009introduction,joos2013decoherence,RevModPhys.75.715,zurek1993preferred}. This can be directly seen through the three-time Leggett-Garg inequality (LGI) \cite{PhysRevLett.54.857,Leggett_2008,10.1007/978-88-470-5217-8_2}, where the authors considered measuring a dichotomous ($\pm 1$ valued) observable $\hat{Q}$ at three different instances $t_1,t_2$ and $t_3$, as the system evolves in time. The quantity of interest here is a linear combination $K_3 :=C_{12} + C_{23} - C_{13}$ of two-time correlators $C_{ij}=\langle q(t_i)q(t_j) \rangle$, where $q(t_k)=\pm 1$ is the  dichotomic outcome of $\hat{Q}$ at time $t_k$ and $\langle . \rangle$ means an average over many runs of the experiment. In a classical macrorealistic system, $K_3$ remains bounded as $-3 \leq K_3 \leq 1$, which is known as  3-term LGI. Quantum systems do not obey classical macrorealism and hence they surpass the upper bound of LGI. Thus the violation of LGI acts as a benchmark of non-classical behavior in terms of non-macrorealism and it is observed experimentally in a variety quantum systems \cite{palacios2010experimental,PhysRevLett.107.130402,PhysRevA.87.052102,PhysRevLett.107.090401}. It is also observed \cite{PhysRevLett.107.090401, PhysRevLett.107.130402} that the violation of LGI dies out with time due to decoherence which indicates the emergence of classicality.

These early experiments \cite{palacios2010experimental,PhysRevLett.107.130402} and theoretical studies \cite{PhysRevLett.111.020403, fritz2010quantum} suggest that for a two-level quantum system (TLS), evolving under unitary dynamics, the maximum violation of $K_3$ is constrained by the Temporal Tsirelson bound (TTB), also known as L\"{u}der's Bound, which reads $\left( K_3\right)_{\rm max} \leq 1.5$. This shows that even quantum theory restricts $\left( K_3\right)_{\rm max}$ way bellow its algebraic maximum value $3$. Incidentally, violation of TTB is reported either by replacing the TLS with a multi-level system \cite{PhysRevLett.113.050401, wang2017enhanced, katiyar2017experimental} or by replacing unitary evolution with a non-Hermitian $\mathcal{PT}$ symmetric evolution \cite{PhysRevA.103.032420, PhysRevA.108.032202, PhysRevA.109.042205,wu2023maximizing}.

Since the superposition of states allows a system to exhibit non-macrorealistic behavior by violating LGI upto TTB, we ask whether it is possible to get enhanced non-macrorealism that transcends TTB  by having superposition not only in the description of state, but also in time evolution unitary!  In this work, we investigate time translation of a TLS by subjecting it to an effective unitary transformation achievable via superposition of unitaries and find $K_3$ violates TTB under such dynamics. In-fact, this enhanced non-macrorealism, as quantified by the violation of LGI beyond TTB, grows with increasing superposition between the unitaries. Furthermore, the unitary superposition provides more robustness  to the TLS against decoherence in a noisy environment by allowing it to violate LGI and thereby preserving its quantum behavior for a much longer time than usual. We employ Nuclear Magnetic Resonance (NMR) architecture for  demonstrating an experimental realization of the superposed unitary operations to verify our theoretical findings. We also provide an intuitive explanation for both the enhanced non-macrorealism and noise robustness  in terms of non-linear speed of evolution (SOE) of the TLS arising from the superposition of unitaries. 

\emph{Superposition of unitaries ---} In the case of quantum states, the well-known superposition between two physically distinct states, say $\Ket{0}$ and $\Ket{1}$ \footnote{Throughout this article, $\hat{\sigma}_{k}$ represents Pauli operator along $\hat{k}$ direction in the Bloch sphere of a TLS , while $\Ket{0}$ and $\Ket{1}$ are eigenstates of $\hat{\sigma}_{z}$ with eigenvalues $1$ and $-1$, respectively}, is formed as $\Ket{\psi} = [c_0 \Ket{0} + c_1  \Ket{1}]/N,~\mathrm{with}~N^2 = \dnorm{\widetilde{\psi}}$, where ${\Ket{\widetilde{\psi}}}:= c_0\Ket{0}+c_1\Ket{1}$ being the unnormalized state.  In the same spirit, we propose to form a superposition between two distinct unitary operators $U_{0}(t_f,t_0)=\exp(-\im \hat{\sigma}_{n}\omega \delta /2)$ and $U_{1}(t_f,t_0)=\exp(-\im \hat{\sigma}_{m}\omega \delta/2)$, the first one causes a rotation about $\hat{n}$ and the second one about $\hat{m}$ in the Bloch sphere in time $\delta = t_f - t_0$ with frequency $\omega$. Considering $\alpha \in [0, \pi/2]$ as the superposition parameter (SP), the resultant unitary is constructed as
\begin{gather}
\mathcal{U}(t_f,t_0) = \frac{\sin \alpha \, U_{0}(t_f,t_0) + \cos \alpha \, U_{1}(t_f,t_0)}{\mathcal{N}(t_f,t_0)}, \label{eq:sup} \\
\mathrm{with}~\mathcal{N}^{2}(t_f,t_0)=\frac{1}{2}\mathrm{tr}\left[\widetilde{\mathcal{U}}(t_f,t_0)\widetilde{\mathcal{U}}^{\dagger}(t_f,t_0)\right],  \label{eq:norm_def}
\end{gather}      
where the unnormalized operator $\widetilde{\mathcal{U}}(t_f,t_0)$ is what sits in the numerator of Eq.~\Eqref{eq:sup}. For the resultant operator $\mathcal{U}(t)$ to be unitary, we need $\widetilde{\mathcal{U}}(t_f,t_0)$ to satisfy $\widetilde{\mathcal{U}}(t_f,t_0)\, \widetilde{\mathcal{U}}^{\dagger}(t_f,t_0) \propto \mathbbm{1}$, which is true as long as $-1 < \hat{n} \cdot \hat{m} \leq 1$. Note $\mathcal{U}(t_f,t_0) \in SU(2)$ and thus it describes rotation of the Bloch vector in Bloch sphere. Also, $\mathcal{U}(t_f,t_0)$ only depends on $\delta = t_f - t_0$, and not on $t_0$.
\begin{figure}
\includegraphics[width=8.6cm, clip = true, trim={4.2cm 7.5cm 3.5cm 0}]
{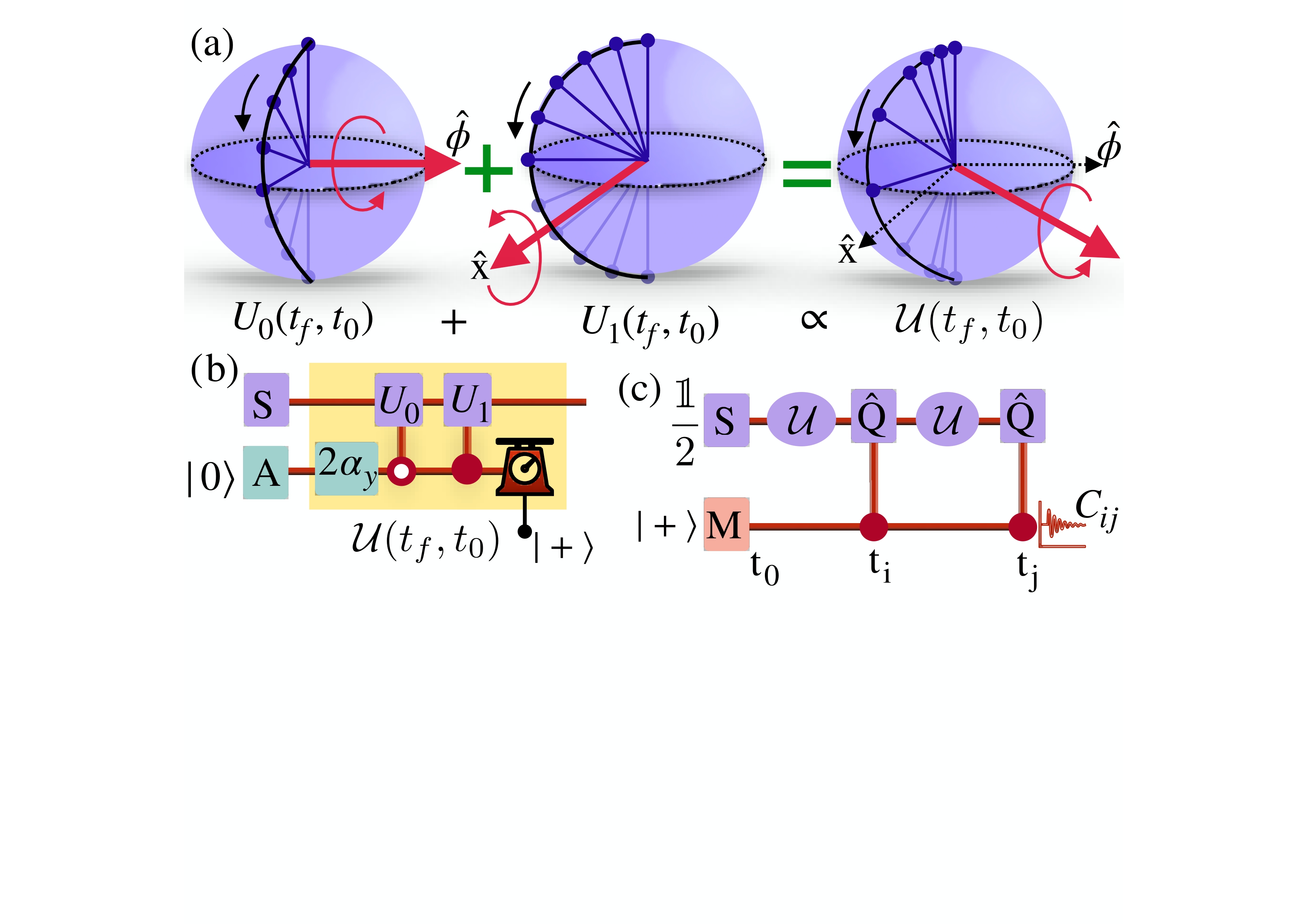}
\caption{(a) The action of a superposition of two unitaries $U_0 = \exp(-\im \hat{\sigma}_\phi \omega \delta/2)$ and $U_1 = \exp(-\im \hat{\sigma}_x \omega \delta/2)$ ($\phi=\pi/2$ here) on the Bloch sphere is two fold : Shifting rotation axis (thick arrow) of the effective unitary $\mathcal{U}$ to lie between $\hat{x}$ and $\hat{\phi}$ and making the SOE under it non-linear in time - as the Bloch vector moving slowly near the poles relatively faster near equator. (b) A quantum circuit for realizing the superposed unitary with SP $\alpha$ by using an ancillary qubit (A). Here $2\alpha_y$ describes a rotation of the Bloch vector by angle $2\alpha$ about $\hat{y}$. (c) The interferometric circuit to determine the two-time correlator $C_{ij}$ corresponding to observable $\hat{Q}$ using an additional qubit M.}
\label{fig:graph_abst}
\end{figure}

To construct the general superposed unitary of Eq.~\Eqref{eq:sup}, consider $\hat{m}=\hat{x}$ and $\hat{n}=\hat{\phi}$, which is a unit vector making an angle $\phi \in [0,\pi)$ with $\hat{x}$
(see Fig.~\ref{fig:graph_abst}~(a)). Taking the initial state of TLS  $\Ket{\psi}_{0}:= \Ket{0}$, we let it transform under the superposed unitary map as $\Ket{\psi}_{t} = \mathcal{U}(t,0)\Ket{0} = \cos(f(t)/2)\Ket{0} - \im \sin(f(t)/2)\exp(\im\theta)\Ket{1}$, where $\theta$ depends on $\alpha$ and $\phi$. Therefore, if we make $t$ to increase continuously from $0$, the map $\mathcal{U}(0,t)$ causes the initial Bloch vector $\hat{z}$ to rotate continuously about an axis $\hat{\theta} = \cos \theta \hat{x} + \sin \theta \hat{y}$. It is crucial to note that this rotation of the Bloch vector happens with a non-linear SOE, which can be defined as $g(t):=\partial_t f(t)$. This $g(t)$ is a non-linear function of $t$ such that the non-linearity increases with increasing SP. Also, the non-linearity in $g(t)$ increases with $\phi$, at a given $\alpha$.  Therefore, if we define the superposed unitary of Eq.~\Eqref{eq:sup} as a transformation from $t_0$ to $t_f$ and make $t_f$ to increase continuously, then the action of superposition is two fold : (i) it shifts the axis of rotation $(\hat{\theta})$ of the Bloch vector to lie in between, and in the same plane of, $\hat{n}$ and $\hat{m}$, and (ii) it makes the SOE of the Bloch vector to be non-linear in $t_f$, such that the non-linearity increases with increasing superposition. For a more formal discussion and mathematical proofs, see Appendix~\ref{append:superpose}.

Since the superposed unitaries of Eq.~\Eqref{eq:sup} do not satisfy the composition law, i.e $\mathcal{U}(t,0) \neq \mathcal{U}(t,\tau)\,\mathcal{U}(\tau,0)$, given $ 0 < \tau < t $, they are not solutions of Schr\"{o}dinger equations. Therefore, to realize them, we use an ancillary qubit (A), initialized in a superposed state $\Ket{\alpha_{\rm{A}}}=\cos \alpha \Ket{0} + \sin \alpha \Ket{1}$ (see Fig.~\ref{fig:graph_abst}~(b)). In the circuit, two controlled gates $U_{T0}(t_f,t_0) = [U_{0}(t_f,t_0)]_{\mathrm{S}} \otimes \outpr{0}{0}_{\mathrm{A}} + \mathbbm{1}_{\mathrm{S}} \otimes \outpr{1}{1}_{\mathrm{A}}$ and $U_{T1}(t_f,t_0) = \mathbbm{1}_{\mathrm{S}} \otimes \outpr{0}{0}_{\mathrm{A}} + [U_{1}(t_f,t_0)]_{\mathrm{S}} \otimes \outpr{1}{1}_{\mathrm{A}}$ are applied back to back. At the end A is post-selected in state $\Ket{+}_{\rm{A}}:=(\Ket{0}+\Ket{1})/\sqrt{2}$, which ensures S to transforms as
\begin{gather}
\rho_{t_f} = \frac{\Bra{+}_{\rm{A}}L(t_f,t_0) \rho_{t_0} \otimes \proj{\alpha_A} L^{\dagger}(t_f,t_0) \Ket{+}_A}{\rm{tr}\left[ \proj{+}_{\mathrm{A}} L(t_f,t_0) \rho_{t_0} \otimes \proj{\alpha_A} L^{\dagger}(t_f,t_0)\right]} \nonumber \\
= \frac{\widetilde{\mathcal{U}}(t_f,t_0) \rho_{t_0} \widetilde{\mathcal{U}}^{\dagger}(t_f,t_0)}{\mathcal{N}(t_f,t_0)} = \mathcal{U}(t_f,t_0) \rho_{t_0} \mathcal{U}^{\dagger}(t_f,t_0), \label{eq:ps}
\end{gather}
where $L(\cdot) = U_{T1}(\cdot)\,U_{T_0}(\cdot)$, and $\rho_t$ describes the state of S at time $t$. Thus, the circuit shown in Fig.~\ref{fig:graph_abst}~(b) describes a physical realization of the system undergoing superposed unitary transformation in time $\delta=t_f-t_0$.

\emph{The three-time LGI on a TLS evolving under  superposed unitary $\mathcal{U}$ ---} Choosing  $\rho_{t_0}=\mathbbm{1}/2$, the two point correlator $C_{ij}$ for dichotomous observable $\hat{Q}$ can be directly determined using the interferometric circuit \cite{PhysRevLett.104.160501} shown in Fig.~\ref{fig:graph_abst}~(c). Here the correlation between measurement outcomes of $\hat{Q}$ is stored in the phase of an additional qubit (M), which is initiated in a coherent state $\Ket{+}_{\rm{M}}$. The measurement outcome of M's coherence $\mathcal{C}:=[ \langle \hat{\sigma}_{x} \rangle + \im \langle \hat{\sigma}_{y} \rangle ]/2$ directly gives $C_{ij}$ as
\begin{equation}
\mathrm{Re} \, [\mathcal{C}] = \frac{1}{2}\mathrm{tr} \left[ \hat{Q}\,\mathcal{U}(t_j,t_i) \hat{Q} \,\mathcal{U}^{\dagger}(t_j,t_i) \right] = C_{ij}.
\label{eq:circ_corr}
\end{equation} 
Note that here the first measurement at time $t_i$ is noninvasive as demanded for an LGI experiment \cite{PhysRevLett.54.857, Leggett_2008,10.1007/978-88-470-5217-8_2}. Also, the actual measurement is performed only at the end of the circuit which ensures no post-selection to happen at time $t_i$. 

\begin{figure}
\includegraphics[clip=true]{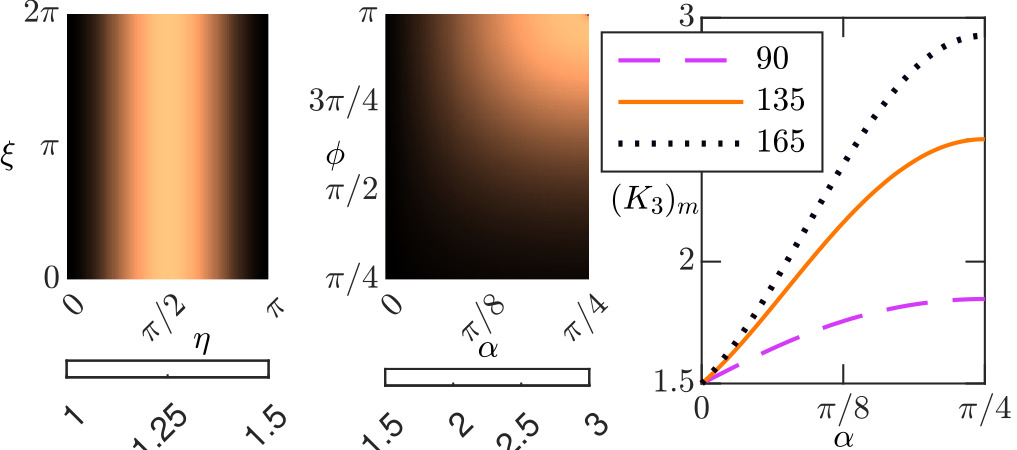} \\
(a) \hspace{2.1cm} (b) \hspace{2.4cm} (c)
\caption{(a) When the system evolves under $U_1$,~ $K^{\rm max}_3$ is evaluated and color plotted against $\xi$ and $\eta$, which are the longitude and co-latitude, respectively, of rotation axis of $U_0$  on the Bloch sphere. The plot shows $K_3$ is upper bounded by TTB=$1.5$. (b) When the system evolves under superposition of unitaries, $K^{\rm max}_3$ is evaluated and color plotted  against $\alpha$ (the amount of superposition) and $\phi$ (the angle between the rotation axis of the unitaries which are being superposed). For each $\phi \in (0,\pi)$, $K^{\rm max}_3$ increases with increaseing $\alpha$. (c) The increase of  $K^{\rm max}_3$ with increase in  SP $\alpha$ is clearly shown for $\phi=90^0$ (dashed line), $\phi = 135^0$ (solid line)  and $\phi = 165^0$ (dotted line) . }
\label{fig:Th_Plots}
\end{figure}
To evaluate $K_3$, we set $\hat{Q}=\hat{\sigma}_z$ and determine the correlators $C_{12}$, $C_{23}$ and $C_{13}$ for the choice of $t_0 = t_1=0,~t_2=t$ and $t_3=2t$. Since $\mathcal{U}(t_f,t_0)$ of Eq.~\Eqref{eq:sup} depends only on $\delta=t_f-t_0$, and not on $t_0$, we get $C_{12}=C_{23}$ and $K_3 = 2C_{12}-C_{13}$. First we consider the zero superposition case of $\alpha = 0$. We align the $\hat{m}$ along an arbitrary direction $ \hat{m}=\sin\eta\cos\xi\,\hat{x}+\sin\eta\sin\xi\,\hat{y}+\cos\eta\,\hat{z} $ in the Bloch sphere and determine  $K^{\rm max}_3$  over one full cycle $\omega t\in [0, 2\pi]$. We repeat this for each possible $\eta \in [0, \pi]$ and $\xi \in [0, 2\pi)$.  The result is shown in Fig.~\ref{fig:Th_Plots}~(a) which displays that the $K^{\rm max}_3$ only depends on $\eta$ and reaches maximum of $3/2$ when $\hat{m}$ lies in the equator ($\eta = \pi/2$). This confirms the existence of TTB as the upper bound of LGI for any choice of $\hat{m}$. For non-zero superposition, we consider the general case by aligning $\hat{m}$ and $\hat{n}$ along $\hat{x}$ and $\hat{\phi}$, and compute the maximum $K^{\rm max}_3$ over one full cycle $\omega t \in [0, 2\pi]$ for full range of parameters $\alpha \in [0, \pi/4]$ and $\phi \in (0, \pi)$. The result is plotted in  Fig.~\ref{fig:Th_Plots}~(b) which shows increase of $K^{\rm max}_3$ beyond TTB with increasing $\alpha$. This growing violation of TTB with rising $\alpha$ is plotted explicitly in Fig.~\ref{fig:Th_Plots}~(c) for $\phi = 90^o, 135^o$ and $165^o$. These results clearly show that for any given $\phi$, the act of superposition in the unitary produces enhanced violation of LGI which increases with increase in the amount of superposition. Moreover, violation of TTB increases with $\phi$ for a fixed $\alpha$ and $K_3$ approaches its algebraic maximum  $K^{\rm max}_3=3$ at $\alpha = \pi/4$ as $\phi \rightarrow \pi$.    

\begin{figure}
\includegraphics[width=8cm, clip=true, trim={3.9cm 3.4cm 4cm 0}]{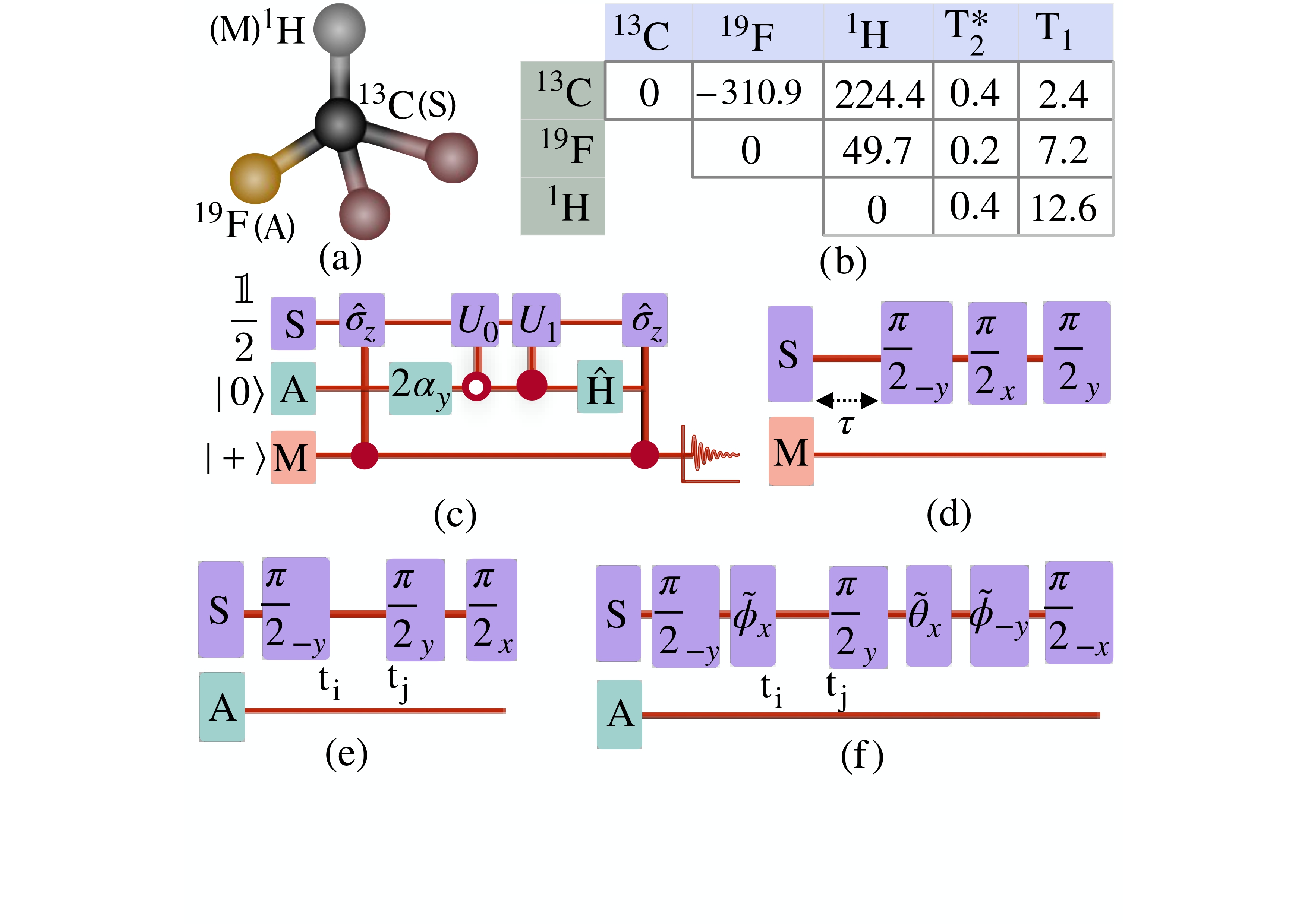}
\caption{(a) The molecular structure of 13C-DBFM with its qubits labelled. (b) Values of the scalar coupling constants in Hz ($J_{ij}$ in $i,j$'th off-diagonal element), resonance offsets (diagonal) along with relaxation time constants ($T_1$ and $T_{2}^{*}$ in s). (c) Quantum circuit for determining $C_{ij}$ of S by measuring the coherence of M in the end. Here $\hat{\rm{H}}$ signifies a Hadamard gate and $2\alpha_y$ signifies rotation in the Bloch sphere by angle $2\alpha$ about $\hat{y}$. (d-f) The quantum circuits used to realize (d) controlled-$\hat{\sigma}_z$ gate (e) $U_{T0}$ and (f) $U_{T1}$, where $\tilde{\phi} = \phi - \pi$ and $\tilde{\theta}=(\pi - t)/2$. Delays represent evolution under the respective scalar couplings and $\tau = 1/2J_{\rm{SM}}$. See Appendix \ref{append:Circuit} for details. }
\label{fig:circuits}
\end{figure}

\emph{Experiments in NMR --- } Consider the three spin-$1/2$ nuclei of 13C-dibromofluoromethane (DBFM) molecule as a three qubit quantum register (see Fig.~\ref{fig:circuits}~(a)). We identify $^{13}$C qubit as our system (S), $^{19}$F qubit as A, and $^{1}$H qubit as M. In a strong magnetic field of $11.7$ T (along $\hat{z}$) inside a Bruker $500$ MHz NMR spectrometer, the liquid ensemble of 13C-DBFM, dissolved in Acetone-D6, rests in thermal equilibrium at an ambient temperature of $300$ K. Under high temperature-high field assumption  \cite{cavanagh1996protein}, the density matrix of the quantum register reads $\rho_{\mathrm{th}} = \mathbbm{1}/8 + \epsilon (\gamma_{\rm{C}} I_z^{\mathrm{S}} + \gamma_{\rm{F}} I_z^{\mathrm{A}} + \gamma_{\rm{H}} I_z^{\mathrm{M}})$, where $\gamma_i$ is the gyro-magnetic ratio of the $i$'th nucleus, $I_{z}^{v}:= \hat{\sigma}^{v}_z /2$, and the purity factor $\epsilon \approx 10^{-5}$.  Using secular approximation in a triply-rotating frame, rotating at the resonant frequency of each nucleus, the three-qubit quantum register's Hamiltonian can be written as \cite{cavanagh1996protein,levitt2008spin}
$\mathcal{H}_{\mathrm{NMR}} = 2\pi \left( J_{\mathrm{SA}}I_z^{\mathrm{S}} I_z^{\mathrm{A}} +  J_{\mathrm{SM}} I_z^{\mathrm{S}} I_z^{\mathrm{M}} +  J_{\mathrm{AM}} I_z^{\mathrm{A}} I_z^{\mathrm{M}} \right)$,
where $J$'s are the respective scalar coupling constants, values of which along with the relaxation time constants of the quantum register are listed in Fig.~\ref{fig:circuits}~(b).

We start the experiment by initializing the quantum register in the state $\rho_{in} = (\mathbbm{1}/2)_{\mathrm{S}} \otimes \proj{0\,+}_{\mathrm{AM}}$ from the thermal state $\rho_{\mathrm{th}}$ via preparation of a pseudo-pure state \cite{cory1997ensemble,gershenfeld1997bulk} between $A$ and $M$ using spatial averaging with spin selective pulses \cite{cory1998nuclear}. After initialization, the quantum circuit of Fig.~\ref{fig:circuits}~(c) is used to perform the experiments, where we combine the circuit of Fig.~\ref{fig:graph_abst}~(b) for realizing the superposition of unitaries with the interferometric circuit of Fig.~\ref{fig:graph_abst}~(c) for measuring the correlator $C_{ij}$. The Hadamard gate $\hat{\rm{H}}$ on A at the end of the circuit makes post-selecting A in the state $\Ket{+}_{\rm{A}}$ equivalent to doing the same in state $\Ket{0}_{\rm{A}}$. A measurement of the real part of the coherence $\mathcal{C}_{\rm{M}}$ at the end of the circuit gives   
\begin{equation}
\rm{Re}\,[\mathcal{C}_{\rm{M}}] = \frac{1}{4} (T_{+}(t_j,t_i) + T_{-}(t_j,t_i)) \label{eqS}
\end{equation}
\begin{equation}
\mathrm{where,}~ T_{\pm}(t_j,t_i)  = \mathrm{tr} \left[ \hat{\sigma}_z \,\widetilde{\mathcal{U}}_{\pm}(t_j,t_i)\, \hat{\sigma}_z \frac{\mathbbm{1}}{2}\, \widetilde{\mathcal{U}}_{\pm}^{\dagger}(t_j,t_i)\right], \nonumber
\end{equation}
with $\widetilde{\mathcal{U}}_{\pm}=\cos(\alpha)U_0(t_j,t_i) \pm \sin(\alpha)U_1(t_j,t_i)$. Interestingly, the terms $T_{+}(t_j,t_i)$ and ${T}_{-}(t_j,t_i)$ can be measured separately via 
distinct spectral lines in NMR, which is equivalent of post selecting A in $\{\Ket{0},\Ket{1}\}$ basis.  (see Appendix \ref{append:Exp} for details).    
\begin{figure}
\includegraphics[clip=true]{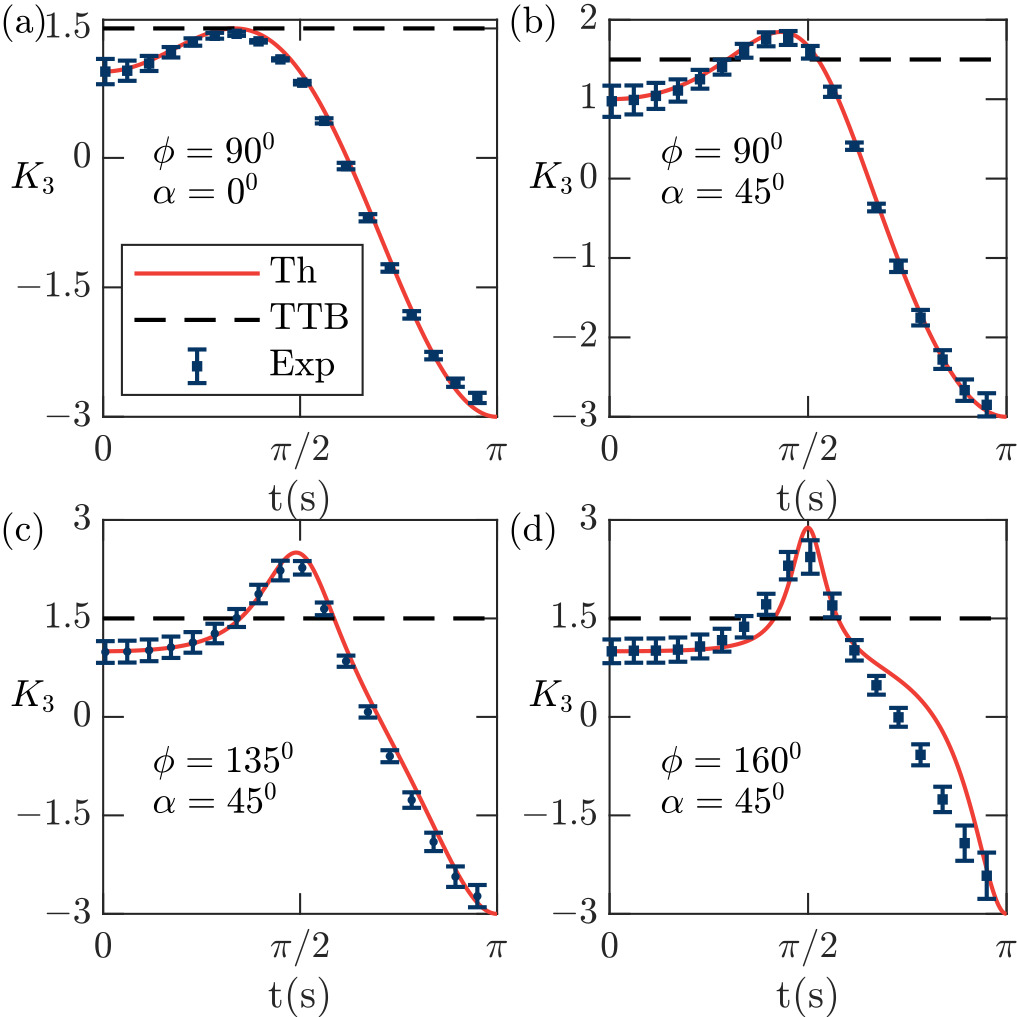}
\caption{Experimentally measured values of $K_3$ (dots with vertical error bars) are plotted along with the their theoretically predicted curves (solid thin curves) for four different values of $\alpha$ and $\phi$. Error bars represent vector sum of systematic (RF in-homogeneity) and random errors (thermal noise).}
\label{fig:Exp}
\end{figure}
Just switching off two controlled $\hat{\sigma}_z$ gates, an identical run of the quantum circuit of fig.~\ref{fig:circuits}~(c) gives the measurement outcome $\rm{Re}\,[\mathcal{C}_{\mathrm{M}}] = \frac{1}{4} (\mathcal{N}(t_j,t_i) + \tilde{\mathcal{N}}(t_j,t_i))$, with  $\mathcal{N}(t_j,t_i)$ as defined in Eq.~\Eqref{eq:norm_def} and $\tilde{\mathcal{N}}(t_j,t_i) = \sqrt{\mathrm{tr}[\mathcal{U}_{-}^{\dagger}(t_j,t_i)\mathcal{U}_{-}(t_j,t_i)}]$. $\mathcal{N}(t_j,t_i)$ and $\tilde{\mathcal{N}}(t_j,t_i)$ can again be measured separately as before. It can be easily seen from Eqs.~\Eqref{eq:sup} and \Eqref{eq:circ_corr} that $C_{ij}=T(t_j,t_i)/\mathcal{N}(t_j,t_i)$. This allows direct experimental measurement of $C_{ij}$ by setting appropriate $t_i$ and $t_j$ in the pulse sequence of Fig.~\ref{fig:circuits}~(e) and (f). Thus, it works as a general method that effectively realizes the superposed unitary channel $\mathcal{U}(t_j,t_i)$ of Eq.~\Eqref{eq:sup}
. Note that our method allows to choose any value of the SP $\alpha$ and align $\hat{n}$ along any direction by setting appropriate value of $\alpha$ and $\phi$ in the pulse sequences of Fig.~\ref{fig:circuits}~(e) and (f).  

We set $t_1 = 0$, $t_2 = t$ and $t_3 = 2t$ as before in our experiments. For no superposition, we set $\alpha=0$ and measure $K_3$ while increasing $\omega t$ from $0$ to $\pi$. Results are shown in Fig.~\ref{fig:Exp}~(a) confirming the maximum $K_3$ is exactly the TTB as predicted. Next, we consider the maximum superposition $\alpha = \pi/4$  at $\phi = \pi/2$ and measure $K_3$ with increasing $\omega t$ and observe a clear violation of TTB (see Fig. \ref{fig:Exp}~(b)). We increase $\phi$ to $135^0$ and $160^0$ at maximum sperposition of $\alpha = \pi/4$ and measure $K_3$ with varying $\omega t$. We observe increasing violation of TTB with increasing $\phi$, as predicted theoretically. Experimentally measured value of  $K^{\rm max}_3$ reads $2.27 \pm 0.1$ for $\phi = 135^0$, showing violation of TTB by more than $7$ times the experimental uncertainty (Fig.~\ref{fig:Exp}~(c)) and $2.43 \pm 0.25$ for $\phi = 160^0$, which violates TTB by more than $3$ times the experimental uncertainty (Fig.~\ref{fig:Exp}~(d)). These results clearly and convincingly demonstrate the enhanced violation of LGI beyond TTB, due to superposition of unitaries.  
\begin{figure}
\includegraphics[clip=true]{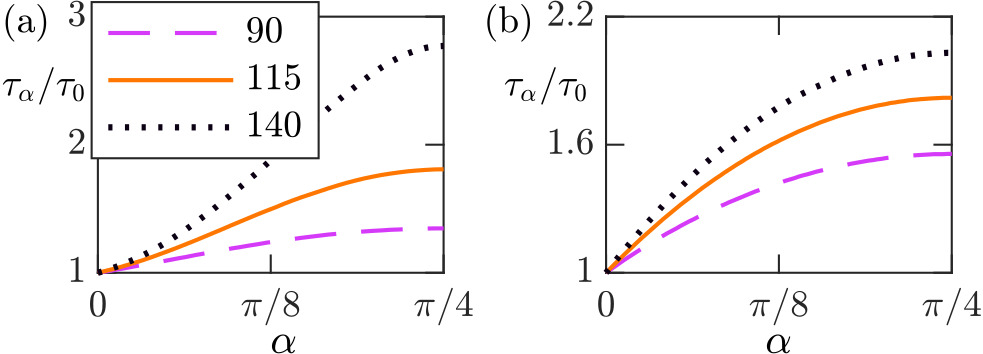} 
\caption{(a) Defining $\tau_{\alpha}$ as the time upto which $K_3 \geq 1$, the gain in robustness $\tau_{\alpha}/\tau_{0}$ is computed by solving Eq.~\Eqref{eq:bloch} and plotted with increasing SP $\alpha$. The plots are showing increase of $\tau_{\alpha}$ with increasing SP for $\phi$ equals $90^{0}$(dashed line), $115^{0}$ (solid line) and $140^{0}$ (dotted line), demonstrating the robustness against decoherence due to the superposition of unitaries. (b) The same is plotted, but using a more modest model of Eq.~\Eqref{eq:lind_blad}, where the additional noise to be encountered in physical realization of the superposition of unitaries are also considered. Results show that the enhanced robustness to persist even in the presence of additional noise. Both computations are done for decay constant $\gamma=1/(4\pi)~\rm{s}^{-1}$.} 
\label{fig:dp}
\end{figure}

We now investigate the effect of markovian dephasing, which is the leading cause of decoherence for most quantum systems, on the dynamics under the superposed unitary. In Eq.~\Eqref{eq:circ_corr}, the observable $\hat{Q}(0)=\hat{\sigma}_z$ is subjected under $\mathcal{U}(t,0)$, with choices $\hat{m}=\hat{x}$ and $\hat{n}=\hat{\phi}$, to get transformed to $\hat{Q}(t)=\mathcal{U}\,(t,0)\hat{\sigma}_{z}\,\mathcal{U}^{\dagger}(t,0)$. Upon increasing $t$ continuously from $0$, this map describes rotation of the Bloch vector $\hat{z}$ about $\hat{\theta}$ with non-linear SOE $g(t)$. Writing the observable at time $t$ as $\hat{Q}(t)=\vec{S}(t) \cdot \vec{\sigma}$, with Bloch vector $\vec{S}(t):=(S_z(t),S_x(t),S_y(t))$ and Pauli vector $\vec{\sigma}:=(\hat{\sigma}_z,\hat{\sigma}_x,\hat{\sigma}_y)$, the rotation is described by the Bloch equation \cite{cavanagh1996protein,roberts1991bloch}, where dephasing can be phenomenologically introduced with decay rate $\gamma$ as 
\begin{equation}
\partial_{t} \vec{S}(t) = g(t)\hat{\theta} \times \vec{S}(t) - \gamma \left( \hat{z} \times \vec{S}(t) \times \hat{z} \right)  . \label{eq:bloch}
\end{equation}

Eq.~\Eqref{eq:bloch} is solved with initial condition $\vec{S}(0)=(1,0,0)$ to get $\vec{S}(t)$ and $\vec{S}(2t)$, from which we directly compute $K_{3}(t) = \hat{z} \cdot (2\vec{S}(t)-\vec{S}(2t))$ using  Eq.~\Eqref{eq:circ_corr}. It is expected that decoherence will eventually cause the system to seize to violate LGI and $K_3$ will fall bellow the classical bound of $1$  as time elapses. The lifetime of the system $\tau_{\alpha}$ is defined as the time up to which $K_3$ remains above $1$ at SP $\alpha$. We compute the gain in lifetime at SP $\alpha$ as $\tau_{\alpha}/\tau_0$ and plot it with increasing SP for different values of $\phi$ in Fig.~\ref{fig:dp}~(a). We find the lifetime of the system to enhance significantly with increasing SP, which shows that the act of superposition provides robustness against such dephasing decay and helps the system to retain its quantum behaviour for longer time.   

Although the above treatment is mathematically correct, in practice the superposed unitary $\mathcal{U}(t,0)$ is realized by adding an ancillary system (A) in a coherent state, letting it evolve jointly with S under an interaction Hamiltonian $\mathcal{H}_{\rm{AS}}$ and finally post-selecting A in $\Ket{+}$ state, as shown in Fig.~\ref{fig:graph_abst}~(b). Therefore, in the physical realization of the superposed unitary, S does not evolve directly with a non-linear SOE as described in Eq.~\Eqref{eq:bloch}. It evolves jointly with A until post-selection, and since A is initialized in a coherent state, dephasing will impact A along with S during the course of this joint evolution. To consider that, we subject the joint state $\rho_{\rm{AS}}$ under a  GKSL master equation \cite{lidar2019lecture,breuer2002theory}
\begin{equation}
\partial_{t} \rho_{\mathrm{AS}}= -\im [\mathcal{H}_{\rm{AS}},\rho_{\mathrm{AS}}] + \frac{\gamma}{2} \sum_{j=\rm{S,A}}(\hat{\sigma}^{j}_{z}\, \rho_{\rm{AS}} \,\hat{\sigma}^{j}_{z} - \rho_{\rm{AS}}), 
\label{eq:lind_blad}
\end{equation}
where $\hat{\sigma}_{z}^{j}$ denotes the action of $\hat{\sigma}_{z}$ on $j$ and identity on the rest, and the interaction Hamiltonian reads $\mathcal{H}_{\rm{AS}}:=\proj{0} \otimes \omega \hat{\sigma}_{n}/2 + \proj{1} \otimes \omega \hat{\sigma}_{m}/2 $. In the absence of decoherence $(\gamma = 0)$, $\mathcal{H}_{\rm{AS}}$ generates the joint evolution $\exp(-\im \mathcal{H}_{\rm{T}}t) = U_{T1}(t,0) \, U_{T0}(t,0) $, which is shown in the circuit of Fig.~\ref{fig:graph_abst}~(b). However, in presence of noise $(\gamma > 0$), and the initial joint state $\rho_{\rm{AS}}(0) =  \proj{\alpha_{\rm{A}}} \otimes \mathbbm{1}_{\rm{S}}/2 $ evolves under Eq.~\Eqref{eq:lind_blad} from $t=0$ to $t$. After which A is post selected in state $\Ket{+}_{\rm{A}}$, allowing S to transform under the map $\mathcal{V}(t,0)$, which describes the effective dynamics under superposed unitary and dephasing. The correlators $C_{12}$ and $C_{13}$ are computed as before with $\mathcal{V}$ replacing $\mathcal{U}$ in Eq.~\Eqref{eq:circ_corr}. From this we compute $K_{3}(t)$ to find the gain in lifetime $\tau_{\alpha}/\tau_{0}$ with increasing SP. Results are plotted for different values of $\phi$ in Fig.~\ref{fig:dp}~(b), which shows that even the presence of additional noise in A in practical realization protocols does not affect the robustness achieved by the superposition of unitaries against dephasing.       
\begin{figure}
\includegraphics[width = 8cm, clip=true]{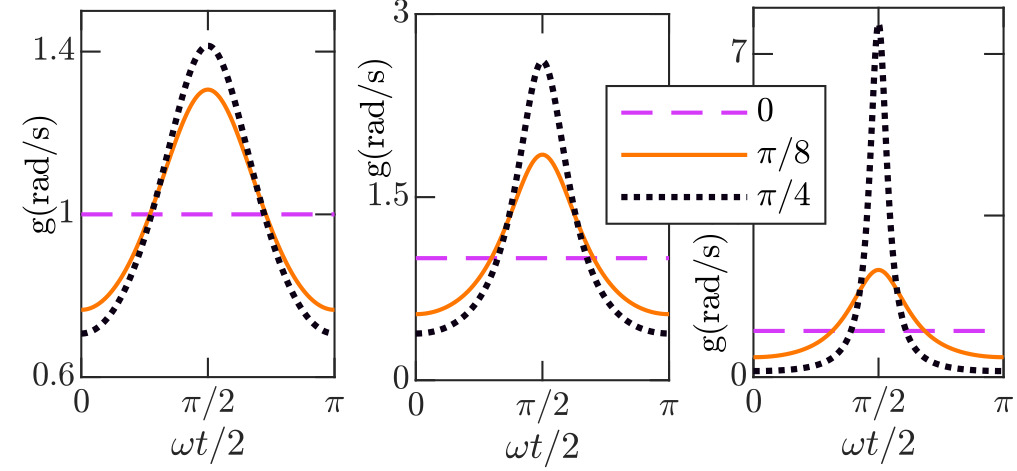} \\
\hspace{0.6cm} (a) \hspace{2cm} (b)\hspace{2cm} (c)
\caption{The SOE of a Bloch vector evolving under the superposition of two unitaries for SP $\alpha=0$ (dashed line), $\pi/8$ (solid line) and $\pi/4$ (dotted line) at (a) $\phi=90^0$, (b) $\phi = 135^0$ and (c) $\phi = 165^0$. At each $\phi$, non-linearity in SOE increases with SP.   }
\label{fig:NL}
\end{figure}

Before concluding, we provide an intuitive way to understand the two main results reported in this article: how the superposition of unitary is causing LGI to go beyond the TTB, and how it is providing the much desired robustness against decoherence. From Eq.~\Eqref{eq:circ_corr}, it can be directly seen that \cite{fritz2010quantum} for $\hat{Q}=\hat{\sigma}_z,~C_{ij} = \mathcal{U}(t_j,t_i)[\hat{z}] \cdot \hat{z}$, where $\hat{z}$ is the initial Bloch vector at time $t_i$, representing the observable, which gets rotated to $\mathcal{U}(t_j,t_i)[\hat{z}]$ at time $t_j$ in Heisenberg picture. At zero superposition, $\mathcal{U}(t_j,t_i)$ describes rotation at frequency $\omega$ about an arbitrary axis. Therefore, for our choice of $t_0=t_1=0,~t_2=t$ and $t_3=2t$, if it rotates $\hat{z}$ by an angle $\omega t/2$ in time $t$, the rotation in time $2t$ will be $\omega t $, resulting $K_3 = 2\cos(\omega t/2)-\cos(2\omega t)$, which always remains upper bounded by $1.5$ for any $\omega t$. This shows that TTB originates from the linearity of rotation of the Bloch vector and necessitates a non-linear SOE for the violation of TTB. As an example, for $K_3$ to approach its algebraic maximum $3$, we need $C_{12}$ and $C_{13}$ to approach $1$ and $-1$, respectively, at same value of $t$. This requires $\mathcal{U}$ to rotate the vector $\hat{z}$ almost none in $0$ to $t$ but by an angle $\pi$ from $t$ to $2t$ which can only be performed with extreme non-linear SOE. We plot the SOE of the Bloch vector $\hat{z}$ at different values of the $\alpha$, for $\phi=90^0,~135^0$ and $165^0$ in Fig.~\ref{fig:NL}~(a),~(b) and (c), respectively. The figure displays the SOE getting increasingly non-linear with increasing SP for each value of $\phi$, which correspondingly results enhanced violation of TTB with increasing $\alpha$. The robustness against dephasing can also be understood with the same argument. The effect of decoherence increases with increasing co-latitude of the Bloch sphere- dephasing has no effect on the poles and does maximum damage on the equator. Due to non-linearity in the SOE, the Bloch vector spends lesser time in the equator and  remains longer near the pole, thereby saving itself from decoherence. This explains the increased robustness of the system with increasing superposition between the unitaries as described by Eq.~\Eqref{eq:bloch}. Also, Fig.~\ref{fig:NL} shows that at a given $\alpha$, the non-linearity in SOE increases with increasing $\phi$, which explains the increasing violation of TTB in Fig.~\ref{fig:Exp} (c-d) and increasing robustness against dephasing in Fig.~\ref{fig:dp} (a-b) at maximum superposition, with increasing $\phi$. 

\emph{Conclusions--} The drastic contrast between the classical and quantum worlds originates primarily from the ability of quantum systems to be in a superposition of physically distinct states. This breach of `macrorealism' as the source of quantum behavior was quantified with a violation of LGI up-to TTB and has been experimentally confirmed in many quantum systems. The question  arises  whether we get a quantum behavior showcasing enhanced non-macrorealism if we apply the superposition principle not only in the description of states but also in the time evolutions. By constructing unitary operators that can be written as superpositions of other unitaries and letting a quantum system evolve under it, we observe that the system exhibits enhanced violation of LGI that surpasses TTB. We quantify this experimentally and show that the violation increases with increasing superposition between the unitaries. Moreover, we found that the act of superposition has application in providing robustness against decoherence by substantially extending the system's lifetime. Both of these enhanced temporal correlations in terms of LGI and robustness against noise can be attributed to the non-linear SOE arising because of the superposition of unitaries.Our work paves the way for research regarding the implications of this new dynamics under the superposed unitaries across different areas of physics.          

\emph{Acknowledgments.--} AC acknowledges Vishal Varma for useful suggestions regarding software setting of the plots. HSK thanks NCN Poland, ChistEra-2023/05/Y/ST2/00005 under the project Modern Device Independent Cryptography (MoDIC). T.S.M. acknowledges funding from DST/ICPS/QuST/2019/Q67 and I-HUB QTF.  ARU is funded by DST/ICPS/QuST/2019/Q107.

\appendix
\section{Properties of the superposed unitaries and defining the SOE}\label{append:superpose}

As mentioned in the main text,  we take two arbitrary directions $\hat{n} = n_x \hat{x} + n_y \hat{y} + n_z \hat{z}$ and $\hat{m} = m_x \hat{x} + m_y \hat{y} + m_z \hat{z}$ in the Bloch sphere. We then construct a superposition between $U_0 (t_f, t_0) = \exp(-\im \hat{\sigma}_n \omega \delta /2)$ and $U_1 (t_f, t_0) = \exp(-\im \hat{\sigma}_m \omega \delta /2)$ according to Eq.~\Eqref{eq:sup} of main text to form $\mathcal{U}(t_f, t_0)$. The unnormalized operator reads 
\begin{gather}
\widetilde{\mathcal{U}}(t_f,t_0) = \sin(\alpha) U_0(t_f,t_0) + \cos(\alpha) U_1(t_f,t_0) \nonumber \\
= \left( \sin(\alpha) + \cos(\alpha) \right) \cos \frac{\omega \delta}{2} \mathbbm{1} \nonumber \\
- \im \left( \sin(\alpha) \hat{\sigma}_n + \cos(\alpha) \hat{\sigma}_m \right) \sin \frac{\omega \delta}{2}.
\end{gather}
Using this, we find
\begin{gather}
\widetilde{\mathcal{U}}(t_f,t_0) \widetilde{\mathcal{U}}^{\dagger}(t_f,t_0) 
= \nonumber \\
\underbrace{ \left[ 1 + \sin(2 \alpha) \left( \cos^{2} \frac{\omega \delta}{2} + \hat{n} \cdot \hat{m} \, \sin^{2} \frac{\omega \delta}{2} \right)\right]}_\text{$\mathcal{N}(t_f,t_0)$}   \mathbbm{1}, 
\end{gather}
which ensures that the superposed operator $\mathcal{U}(t_f,t_0) =\widetilde{\mathcal{U}}(t_f,t_0) / \sqrt{\mathcal{N}(t_f,t_0)} $ is always an unitary operator for arbitrary values of $t_f$ and $t_0$ and for arbitrary choices of directions $\hat{n},~\hat{m}$, as long as $-1 < \hat{n} \cdot \hat{m} \leq 1$, which is required to ensure the positivity of $\mathcal{N}(t_f,t_0)$ for all values of $t_0,~t_f$ and $\alpha \in [0,\pi/4]$.

Without loss of generality, we consider $\hat{m}:=\hat{x}$ and $\hat{n}:=\hat{\phi}$ as in the main text, where $\hat{\phi}$ is a unit vector making an angle $\phi \in [0,\pi)$  with $\hat{x}$. Taking, $t_0=0$ and $t_f=t$, the superposed unitary can be expanded as 
\begin{widetext}
\begin{gather*}
\mathcal{U}(t,0) = \frac{\begin{bmatrix}
\left(\cos \alpha + \sin \alpha \right) \cos \frac{\omega t}{2} & -\im \left(\cos \alpha + \expo{-\im \phi} \sin \alpha \right) \sin \frac{\omega t}{2}  \\
-\im \left(\cos \alpha + \expo{\im \phi} \sin \alpha \right) \sin \frac{\omega t}{2} &
\left( \cos \alpha + \sin \alpha \right) \cos \frac{\omega t}{2}
\end{bmatrix}}{\sqrt{1+\sin(2\alpha)[\cos^{2}\frac{\omega t}{2} + \cos\phi \sin^{2}\frac{\omega t}{2}]}}   
= \underbrace{\frac{\left(\cos \alpha + \sin \alpha \right) \cos \frac{\omega t}{2}}{\sqrt{1+\sin(2\alpha)[\cos^{2}\frac{\omega t}{2} + \cos\phi \sin^{2}\frac{\omega t}{2}]}}}_\text{$\cos [f( t)/2]$} \mathbbm{1} \\
- \im  \underbrace{\frac{\sqrt{1 + \cos \phi \sin (2\alpha)}\sin \frac{\omega t}{2}}{\sqrt{1+\sin(2\alpha)[\cos^{2}\frac{\omega t}{2} + \cos\phi \sin^{2}\frac{\omega t}{2}]}}}_\text{$\sin[f( t)/2]$} \left( \underbrace{\frac{\cos \alpha + \cos \phi \sin \alpha}{\sqrt{1 + \cos \phi \sin (2\alpha)}}}_\text{$\cos \theta$} \hat{\sigma}_x + \underbrace{\frac{\sin \alpha \sin \phi}{\sqrt{1 + \cos \phi \sin (2\alpha)}}}_\text{$\sin \theta$} \hat{\sigma}_y \right) 
= \cos \left[\frac{f(t)}{2}\right] \mathbbm{1} - \im \sin \left[\frac{f(t)}{2}\right] \hat{\sigma}_{\mathrm{SP}},
\label{eq:sup_detailed}
\end{gather*}
\end{widetext}
where $\hat{\sigma}_{\rm{SP}}:=\cos \theta \hat{\sigma}_x + \sin \theta \hat{\sigma}_y$. To define the speed of evolution (SOE) $g(t)$, we take the initial state $\Ket{\psi_0}:=\Ket{0}$ and let it transform under the superposed unitary as 
\begin{gather}
\Ket{\psi_t} = \mathcal{U}(t,0)\Ket{0} = \cos \left[\frac{f(t)}{2}\right] \Ket{0} -\im \sin \left[\frac{f(t)}{2}\right] \expo{\im \theta} \Ket{1} \nonumber \\
= \cos \left[\frac{f(t)}{2}\right] \Ket{0} + \sin \left[\frac{f(t)}{2}\right] \expo{\im (\theta - \pi/2)} \Ket{1}. \label{eq:rot_soe}
\end{gather}
Therefore, if we increase $t$ continuously from $0$, Eq.~\Eqref{eq:rot_soe} describes counter-clockwise rotation of the Bloch vector $\hat{z}$ about axis $\hat{\theta}:=\cos \theta \hat{x} + \sin \theta \hat{y}$ in the Bloch Sphere. SOE $(g(t))$ can now be defined as the rate of change of the co-latitude (which is the only degree of freedom that is changing here) of the Bloch vector with $t$ as
\begin{align}
g(t) =  \partial_t f(t) =  \frac{-2\partial_t \cos [f(t)/2]}{\sin[f(t)/2]}.
\end{align}
Computing the derivatives,  we get
\begin{gather}
\partial_t \cos [f(t)] \nonumber \\ = - \frac{\omega \left(1 + \sin (2\alpha) \cos \phi \right) \left( \cos \alpha + \sin \alpha \right) \sin \frac{\omega t}{2} }{ \left[1+\sin(2\alpha)(\cos^{2} \frac{\omega t}{2} + \cos \phi \sin^{2} \frac{\omega t}{2}) \right]^{3/2}},
\end{gather}
which directly gives a closed form expression for the SOE. Of course, SOE can be defined more generally by taking a generic initial state as $\Ket{0}$ instead of $\Ket{0}$. In that case, both the co-latitude and longitude of the Bloch vector will change with $t$, and therefore $g(t)$ will read as the magnitude of the vector sum of these two rate of changes. However, the special case presented here suffices the requirements of this paper.

%\section{Derivation of the two time Correlator} 
%\label{append:K3} 
%Here we derive the expression of the two time correlator $C_{ij}$ as given in Eq.~\Eqref{eq:corr_th} in the main text. We consider the dichotomic observable to be $Q$ with outcomes $m_a$ and $m_b$ and the initial state of the system is assumed to be $\rho_0$. Since $[Q,\rho_0]=0$, we express both of them in the same eigen-basis as $\rho_0 = P_a \proj{a} + P_b \proj{b}$ and $Q = m_a \proj{a} + m_b \proj{b}$. We also consider $V(t_i,t_j) \proj{a} V^{\dagger}(t_i,t_j) := \rho_a$ and $V(t_i,t_j) \proj{b} V^{\dagger}(t_i,t_j) := \rho_b$ .  Now we can write 
%\begin{gather}
%C_{ij} = m_{a}^{2}P_a \langle a | \rho_a | a \rangle + m_{b}^2 P_b \langle b | \rho_b | b \rangle \nonumber \\ 
%+ m_{a}m_{b} ( P_a \langle b | \rho_a | b \rangle + \langle a | \rho_b | a \rangle ) \nonumber \\
%= m_{a} (\langle a | V(t_i,t_j) (m_a P_a \proj{a} + m_b P_b \proj{b}) V^{\dagger}(t_i,t_j) | a \rangle) \nonumber \\
%+  m_{b} (\langle b | V(t_i,t_j) (m_a P_a \proj{a} + m_b P_b \proj{b}) V^{\dagger}(t_i,t_j) | b \rangle) \nonumber \\
%= m_a  \langle a | V(t_i,t_j) Q \rho_0 V^{\dagger}(t_i,t_j) | a \rangle \nonumber \\
% +m_b  \langle b | V(t_i,t_j) Q \rho_0 V^{\dagger}(t_i,t_j) | b \rangle \nonumber \\
%= \mathrm{tr}[Q V(t_i,t_j) Q \rho_0 V^{\dagger}(t_i,t_j)].
%\end{gather}
%This completes the derivation. 
\section{Detailed Derivations of the Quantum Circuits Used}
\label{append:Circuit}
Here we derive the pulse sequence for the controlled gates used in the quantum circuit shown in Figure \ref{fig:circuits}~(d-f) in the main text.
We denote $\proj{0}:=\sigma_0$ and $\proj{1}:=\sigma_1$.
For the controlled $\hat{\sigma}_z$ gate, note
\begin{gather}
U_{cz} = \mathbbm{1}_{S} \otimes \proj{0}_{M} + \hat{\sigma}_{z}^S \otimes \proj{1}_{M} \nonumber \\
= \sigma_{0}^{M} + \sigma_{z}^{S} \sigma_{1}^{M} 
= \expo{\im \sigma_{z}^{S} \sigma_{1}^{M} \pi/2} 
= \expo{\im \sigma_{z}^{S}\pi/4} \expo{-\im \sigma_{z}^{S} \sigma_{z}^{M} \pi/4} \nonumber \\
= \expo{-\im \sigma_{y}^{S}\pi/4} \expo{-\im \sigma_{x}^{S} \pi/4} \expo{\im \sigma_{y}^{S} \pi/4} \expo{-\im \sigma_{z}^{S} \sigma_{z}^{M} \pi/4 },
\end{gather}
which is the circuit shown in Fig.~\ref{fig:circuits}~(d) in the main text.

Now, we note 
\begin{gather}
U_{T0} = (\expo{-\im \sigma_x \omega t/2})_S \otimes \proj{0}_A + \mathbbm{1}_S \otimes \proj{1}_A \nonumber \\
= \sigma_{1}^{A} + \cos(\omega t/2) \sigma_{0}^{A} - \im \sin(\omega t/2) \sigma_{x}^S \sigma_{0}^{A} \nonumber \\
= \expo{-\im \sigma_{0}^{A} \sigma_{x}^{A} \omega t/2} 
= \expo{-\im \sigma_{x}^{S}(1+\sigma_{z}^{A})\omega t/4} \nonumber \\
= \expo{-\im \sigma_{x}^{S}\omega t/4} \expo{-\im \sigma_{y}^{S} \pi/4} \expo{-\im \sigma_{z}^S \sigma_{z}^A \omega t/4} \expo{\im \sigma_{y}^S \pi/4},
\end{gather}
which is the pulse sequence shown in Fig \ref{fig:circuits}~(e) in the main text. By a similar line of reasoning, we find
\begin{gather}
U_{T1} = (\expo{-\im \sigma_{\phi} \omega t/2})_S \otimes \proj{1}_A + \mathbbm{1}_S \otimes \proj{0}_A \nonumber \\
= \sigma_{0}^{A} + \cos(\omega t/2) \sigma_{1}^{A} - \im \sin(\omega t/2) \sigma_{\phi}^S \sigma_{0}^{A} \nonumber \\
= \expo{-\im \sigma_{1}^{A} \sigma_{\phi}^{S}\omega t/2} 
=  \expo{-\im \sigma_{\phi}^{S}(1-\sigma_{z}^{A})\omega t/4} \nonumber \\
= \expo{-\im \sigma_{\phi}^{S}\omega t/4} \expo{\im \sigma_{z}^{S}(\pi-\phi)/2} \expo{-\im \sigma_{x}^{S}\sigma_{z}^{A}\omega \omega t/4} \expo{-\im \sigma_{z}^{S}(\pi - \phi)/2} \nonumber \\
= \expo{\im \sigma_{z}^{S}(\pi-\phi)/2} \expo{\im \sigma_{x}^{S}\omega t/4} \expo{-\im \sigma_{y}^{S}\pi/4} \expo{-\im \sigma_{z}^{S}\sigma_{z}^{A}\omega t/4} \expo{\im \sigma_{y}^{S}\pi/4} \expo{-\im \sigma_{z}^{S}(\pi - \phi)/2} \nonumber \\
= \expo{\im \sigma_{x}^{S}\pi/4} \expo{-\im \sigma_{y}(\pi-\phi)/2} \expo{\im \sigma_{x}^{S}(\omega t-\pi)/4} \expo{-\im \sigma_{y}^{S}\pi/4} \expo{-\im \sigma_{z}^{S}\sigma_{z}^{A}\omega t/4} \nonumber \\
\expo{\im \sigma_{x}^{S}(\pi-\phi)/2} \expo{\im \sigma_{y}^{S}\pi/4},
\end{gather}
which is the pulse sequence shown in Fig \ref{fig:circuits}~(f) in the main text. 

\section{Experimental Details}
\label{append:Exp}
Here we first derive the Eq.~\Eqref{eqS} of the main text. Evolution of the quantum register under the quantum circuit shown in Fig \ref{fig:circuits}~(c) in the main text reads :
\begin{gather}
\frac{1}{2} \left(\proj{0}+\proj{1}+\outpr{0}{1}+\outpr{1}{0}\right)_{M} \nonumber \\
\otimes  (\cos^{2}(\alpha)\proj{0}+\sin^{2}(\alpha)\proj{1}+\cos(\alpha)\sin(\alpha)(\outpr{0}{1}+ \nonumber \\
\outpr{1}{0}))_{A} \otimes \frac{\mathbbm{1}}{2}_{S} \nonumber \\ 	
\downarrow  [U_{T0} U_{T1}]_{AS} \nonumber \\
\frac{1}{2}(... + \outpr{0}{1} + \outpr{1}{0})_{M} \otimes (\cos^{2}(\alpha) \proj{0}_{A} \otimes [U_{T0} \frac{\mathbbm{1}}{2} U^{\dagger}_{T0}]_{S} \nonumber \\
+ \sin^{2}(\alpha) \proj{1}_{A} \otimes [U_{T1} \frac{\mathbbm{1}}{2} U^{\dagger}_{T1}]_{S} + \cos(\alpha) \sin(\alpha) \nonumber \\
\outpr{0}{1}_{A} \otimes [U_{T0} \frac{\mathbbm{1}}{2} U^{\dagger}_{T1}]_{S} + \outpr{1}{0}_{A} \otimes [U_{T1} \frac{\mathbbm{1}}{2} U^{\dagger}_{T0}]_{S}) \nonumber \\
\downarrow \hat{H}_{A} \rightarrow \mathrm{tr}_{AS}[.] \nonumber \\
\rho_{M} = \frac{1}{2}\proj{+} (T_{+} + T_{-}), \label{aeq:T}
\end{gather}
which results the (real) NMR signal $S = \mathrm{tr}[\sigma_{x}\rho_{M}] = \frac{1}{4}(T_{+} +T_{-})$ when $M$ is measured. This completes the derivation of Eq.~\Eqref{eqS}
in the main text. Just following these line of steps, the similar results for $\mathcal{N}$ of the main text can also be derived directly.

Finally, note that during NMR signal acquisition, the final three qubit state effectively evolves under $\mathcal{H}_{\rm{NMR}}$ of the main text. This causes the $T_{+}$ and $T_{-}$ of Eq.~\Eqref{aeq:T} to evolve at different frequencies. Therefore by taking a Fourier transform of the time domain signal we go to the frequency domain where we find $T_{+}$ and $T_{-}$ to be separated. This allows us to post-select the $T$ term.

\bibliography{bibliography}
\end{document}